\documentclass{article}
\usepackage{hiph-preprint}
\usepackage{mathtext}
\usepackage{graphicx}
\volnumber{22} \issuenumber{1} \edyear{2005}                             
\frompage{000} \topage{000}                                              
\recrevdate{22 November 2005}                                            
\def\rightmark{High $p_{t}$ azimuthal correlations - CERES}
\def\leftmark{M. P{\l }osko\'n}

\title{Two-particle azimuthal correlations at high transverse momentum in Pb-Au at $158$~AGeV/c} 
\authors{ 
{Mateusz P{\l }osko\'n$^1$ for the CERES Collaboration 
\index{One, M. P{\l }osko{\'n}} 
}\\[2.812mm]
{\normalsize
\hspace*{-8pt}
$^1$ Institut f\"ur Kernphysik, Universit\"at Frankfurt, Germany\\[0.2ex] 
}}
 
\abstract{The study of two-particle azimuthal correlations at high transverse
momentum has become an important tool to investigate the interaction of hard
partons with the medium formed in high-energy nucleus-nucleus collisions. At
SPS energies, pioneering studies by the CERES collaboration \cite{JANA} indicated a
significant modification of the away-side structure in central collisions. 
Here we present new results emerging from the analysis of the year 2000 data set
recorded with the CERES Time-Projection Chamber, which provides excellent
tracking efficiency and significantly improved momentum determination.}

 
\makeindex
\begin{document}
 
\maketitle

\section{Introduction}

The strong suppression of the yield of hadrons with large transverse momenta in
central Au+Au collisions at RHIC, compared to yields in p+p collisions scaled by the number
of binary nucleon-nucleon collisions has been regarded as a manifestation
of the extremely dense color charged medium created in nucleus-nucleus collisions \cite{STAR,PHENIXdense}. 
It has been predicted that such a dense medium would strongly enhance the energy loss
of a hard-scattered parton and modify its fragmentation function, resulting in the
suppression of high-$p_{t}$ hadrons in the final state. The tool to verify these
expectations is the study of jets and their modifications when traversing the dense
medium. For typical jet energies at SPS and RHIC, a direct reconstruction of jets
is hampered by the huge background of uncorrelated soft particles. On the other
hand, a statistical analysis of the azimuthal angular correlation with respect
to a leading hadron is suitable to verify the jet and di-jet production and address
the question of their in-medium modification \cite{PHENIX}. Due to its full azimuthal coverage
at mid-rapidity the CERES spectrometer at the CERN-SPS is ideally suited for angular 
correlation studies. Here, we present leading hadron azimuthal correlations in Pb-Au 
collisions at $158$~AGeV/c.

\section{Experiment and data analysis}

The present study is based on the analysis of 30 million Pb-Au events at $158$~AGeV/c
recorded with the CERES spectrometer at SPS in the year 2000. The momenta of the charged
particles are determined by the curvatures of their tracks in the radial Time Projection Chamber
(TPC). The achieved momentum resolution is $\Delta p / p = ((2\%)^2 + (1\% p ($~GeV/c~$)^2)^{1/2}$ \cite{MISKO}. 
The detector acceptance is $2\pi$ in azimuth and $2.1$ to $2.7$ in pseudo-rapidity. 
Additionally, due to the limited two-track resolution a separation cut of $\Delta \theta < 10$~mrad has been applied. 
The centrality and the number of participating nucleons have
been estimated with the nuclear overlap model calculation \cite{OVERLAP} using the charged particle multiplicities in the Silicon Drift Detectors
(SDD) and TPC. The data set has been subdivided into three centrality samples:
the $0-5\%$, $5-10\%$ and $10-20\%$ most central events.

For the analysis of leading hadron correlations, we follow a scheme recently
presented by the PHENIX collaboration \cite{PHENIX}. We select trigger particles in the range
$2.5$~GeV/c~$< p_t^{trigger} < 4.0$~GeV/c and calculate the azimuthal angular difference
$\Delta \phi^{same} $ with respect to associated particles with $1.0$~GeV/c~$< p_t^{assoc.} < 2.5$~GeV/c in the same event. 
The normalized yield $Y(\Delta \phi^{same})$ has been divided by the normalized
yield $Y(\Delta \phi^{mixed})$ obtained from mixed event pairs to form a correlation
function:
\begin{equation}\label{correldef}
C(\Delta \phi) = \frac{Y_{same}(\Delta \phi)}{Y_{mixed}(\Delta \phi)} \cdot \frac{\int Y^{mixed}}{\int Y^{same}} 
\end{equation}

\begin{figure}[htb]
\begin{center}
\includegraphics[scale=0.54]{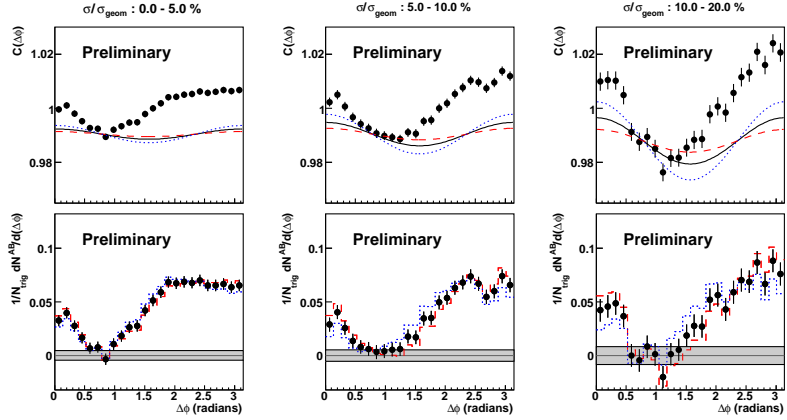}
\caption{\underline{\it Upper row:} Correlation functions for the three centrality bins. The full lines indicate
flow contributions estimated with the ZYAM method. The dashed and dotted
lines represent the flow contributions with ($v^A_2$) and ($v^B_2$) 
varied according to their statistical uncertainties. \underline{\it Lower row:} Conditional yields of the jet-pair 
distributions for the three centrality bins, normalized to the number of triggers. The full, horizontal bands represent
the systematic uncertainty resulting from the flow constant $b_0$ estimation by the ZYAM method. The dashed and dotted lines 
represent distributions that would result from varying ($v^A_2$) and ($v^B_2$) values within their statistical uncertainties.}
\label{fig1}
\end{center}
\end{figure}

The resulting correlation functions are shown in Fig. \ref{fig1}. They exhibit a narrow near-side
structure at $\Delta \phi \approx 0$ and a broad away-side correlation at $\Delta \phi \approx \pi$. 
The back-to-back emission pattern may be attributed to (di-)jet correlations, however, the two-particle
$\Delta \phi$ distributions also contain correlations due to elliptic flow. 
The non-zero two-particle flow contribution ($1 + 2 < v_2^A v_2^B > \cos(2 \Delta \phi)$)
has been approximated by ($ 1 + 2 < v_2^A > < v_2^B > \cos(2 \Delta \phi) $). The flow values 
($v_2^A$) for the trigger and ($v_2^B$) for associated particles
 have been obtained from the reaction plane method \cite{JANA,JOVAN}. The resulting flow
contributions are shown as solid lines in Fig. \ref{fig1}. In order to extract the (di-)jet signal,
the correlation function is decomposed into two contributions: one proportional to
the distribution of the background pairs (containing flow), and the second $J( \Delta \phi)$
representing the (di-)jet pairs:
\begin{equation}\label{correldecomp}
C(\Delta \phi) = b_{0} \cdot (1 + 2 \cdot < v^{A}_{2} > < v^{B}_{2} > \cos(2\Delta \phi)) + J(\Delta \phi).
\end{equation}
The parameter $b_0$ is chosen to match the the Zero Yield At Minimum
(ZYAM) condition \cite{PHENIX}. After the flow contribution is
subtracted from the correlation function, the fully corrected (di-)jet pair
distribution $dN^{AB}_{(Di-)Jet} / d\Delta \phi$  can be constructed. 
The conditional yield distribution of jet-associated particles per trigger is given by:
\begin{equation}\label{condyield}
\frac{1}{N_{trig}}\frac{dN^{AB}}{d\Delta \phi} = \frac{J(\Delta \phi)}{\int(C(\Delta \phi ')d(\Delta \phi '))} \frac{N^{AB}}{N^{A}},
\end{equation}
where $N^A$ is the number of triggers and $N^{AB}$ the total number of $AB$ pairs in
the event sample. The conditional yields have been corrected for the single hadron
efficiency estimated to be $85\%$ at laboratory momenta greater than $1.0$~GeV/c. This
correction leads to a $1.5\%$ systematic uncertainty on the absolute normalization.

\section{Results and conclusions}

\begin{figure}[htb]
\begin{center}
\includegraphics[scale=0.2]{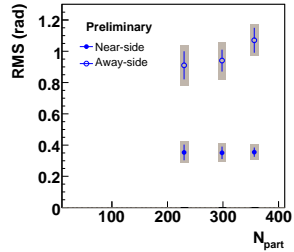}
\caption{Near- and away-side widths (RMS) extracted from the conditional yields.
Statistical (error bars) and systematical uncertainties (filled areas) shown accordingly.}
\label{fig2}
\end{center}
\end{figure}

The conditional yields of the (di-)jet associated particles are shown in Fig. \ref{fig1}. It has been verified that the shapes and the yields of the near-side peaks are consistent with PYTHIA simulations within the systematic uncertainties. On the other hand, it is observed that the away-side structures have a pronounced non-Gaussian shape as centrality increases. 
This significant modification of the back-to-back structure, is possibly caused
by substantial re-interaction of the scattered partons in the medium. In the most
central selection the away-side structure is rather a plateau, revealing the existence
of a local minimum at $\pi$. To extract the widths of the near-side and away-side structures 
we subdivide the $\Delta \phi$ distribution into two parts separated by the global minimum.
The resulting RMS values are shown as function of
the number of participating nucleons in Fig. \ref{fig2}. The widths show no significant centrality
dependence within the range under study and are very similar to the widths observed by the PHENIX collaboration
at RHIC \cite{PHENIX}.

In summary, we have presented leading hadron azimuthal correlations obtained
with the CERES spectrometer. The present findings at top SPS energy show features
which are qualitatively similar to results obtained by the PHENIX collaboration
from Au+Au $\sqrt{s} = 200$~GeV at RHIC \cite{PHENIX}. These observations had been
discussed in the context of conical flow patterns emerging from supersonic partons
in a thermalized colored medium \cite{STOECKER,SHURYAK}. It is remarkable that such mechanisms may
also be at work at SPS energies. However, more detailed studies are required to
substantiate these findings.

\vfill\eject
\end{document}